\documentclass{article}

\usepackage{arxiv}

\usepackage[utf8]{inputenc} % allow utf-8 input
\usepackage[T1]{fontenc}    % use 8-bit T1 fonts
\usepackage{hyperref}       % hyperlinks
\usepackage{url}            % simple URL typesetting
\usepackage{booktabs}       % professional-quality tables
\usepackage{amsfonts}       % blackboard math symbols
\usepackage{nicefrac}       % compact symbols for 1/2, etc.
\usepackage{microtype}      % microtypography
\usepackage{lipsum}
\usepackage{graphicx}
\graphicspath{ {./images/} }

\usepackage{cite}
\usepackage{amsmath,amssymb,amsfonts}
\usepackage{graphicx}
\usepackage{textcomp}
\usepackage{algorithm}
\usepackage{hyperref}
\usepackage{algpseudocode}
\usepackage{pgfplots}
\usepackage{amsmath}
\usepackage{booktabs}
\usepackage{xcolor}
\usepackage{url}
\usepackage{listings}
\usepackage{xcolor}
\usepackage{tabularx}
\usepackage{booktabs} 
\lstdefinelanguage{Cypher}{
  morekeywords={MATCH, WHERE, OPTIONAL, WITH, CASE, WHEN, THEN, ELSE, END, RETURN, CALL, COLLECT, AS, ORDER, BY, LIMIT},
  sensitive=true,
  morecomment=[l]{//},
  morestring=[b]",
}

\def\BibTeX{{\rm B\kern-.05em{\sc i\kern-.025em b}\kern-.08em
    T\kern-.1667em\lower.7ex\hbox{E}\kern-.125emX}}

\title{GraphRAG-Causal: A Novel Graph-Augmented Framework for Causal Reasoning and Annotation in News}

\author{
 Abdul Haque \\
  Dept. of Artificial Intelligence \& Data Science\\
  FAST School of Computing \\
  NUCES Islamabad, Pakisan \\
  \texttt{i211769@nu.edu.pk} \\
   \And
 Umm e Hani \\
  Dept. of Artificial Intelligence \& Data Science\\
  FAST School of Computing \\
  NUCES Islamabad, Pakisan \\
  \texttt{i211715@nu.edu.pk} \\
  \And
 Ahmad Din \\
  Dept. of Artificial Intelligence \& Data Science\\
  FAST School of Computing\\
  NUCES Islamabad, Pakisan \\
  \texttt{ahmad.din@nu.edu.pk} \\
  \And
  Muhammad Babar \\
  Dept. of Computing and Electronics Engineering \\
  Middle East College \\
  Muscat, Oman \\
  \texttt{babarm@mec.edu.om}\\
  \And
  Ali Abbas\\
   Dept. of Computing and Electronics Engineering \\
  Middle East College \\
  Muscat, Oman \\
  \texttt{Aabbas@mec.edu.om}\\
  \And
   Insaf Ullah\\
   Institute for Analytics and Data Science\\
   University of Essex\\
   Colchester, CO4 3SQ, UK\\
   \texttt{Insaf.ullah@essex.ac.uk}
}
\date{}

\begin{document}
\maketitle
\begin{abstract}
GraphRAG-Causal introduces an innovative framework that combines graph-based retrieval with large language models to enhance causal reasoning in news analysis. Traditional NLP approaches often struggle with identifying complex, implicit causal links, especially in low-data scenarios. Our approach addresses these challenges by transforming annotated news headlines into structured causal knowledge graphs. It then employs a hybrid retrieval system that merges semantic embeddings with graph-based structural cues leveraging Neo4j to accurately match and retrieve relevant events. The framework is built on a three-stage pipeline: First, during Data Preparation, news sentences are meticulously annotated and converted into causal graphs capturing cause, effect, and trigger relationships. Next, the Graph Retrieval stage stores these graphs along with their embeddings in a Neo4j database and utilizes hybrid Cypher queries to efficiently identify events that share both semantic and structural similarities with a given query. Finally, the LLM Inference stage utilizes these retrieved causal graphs in a few-shot learning setup with XML-based prompting, enabling robust classification and tagging of causal relationships. Experimental evaluations demonstrate that GraphRAG-Causal achieves an impressive F1-score of 82.1\% on causal classification using just 20 few-shot examples. This approach significantly boosts accuracy and consistency, making it highly suitable for real-time applications in news reliability assessment, misinformation detection, and policy analysis.
\end{abstract}


\section{Introduction}  
The considerable growth of digital news has emphasized the need for advanced methods to precisely extract and interpret the complex causal relationships that underpin headline content. In spite of significant progress through conventional NLP techniques such as RoBERTa \cite{bhatia}, BERT \cite{tanNewsCorpus} and other large language models, these methods are persistent to linguistic signals and fall short in capturing nuanced, multi-causal and nested causal relationships in real-world news data. GraphRAG-Causal introduces a graph-expansion framework that methodologically captures complex nested cause-effect relationships and detection of causal news, an ability which traditional sentence level LLMs struggle with. Our approach instructs explicitly to output in JSON format (keys: Causal tagged sentence, Original sentence, Causal Label), this overcomes the common challenge of consistency and maintainability of LLMs, ensuring easier integration with other news applications. These contributions pave the way for robust and scalable causal inference in real-world news datasets with easier integration.
 
Directly this approach requires a minimal set of labelled examples to train a model that dynamically retrieves relevant graphs from pre-constructed causal news graphs stored inside a graph database. This dual strategy allows the LLM to adapt to new news scenarios and precisely identify even the most implicitly embedded cause-effect relationships. Other methods rely on more annotated data, while our’s performs on par with limited data and less compute power as compared to fully fine tuning an LLM \cite{tanNewsCorpus}. This effectively bridges the gap between sparse supervised learning and complex causal inference, ensuring robust performance in real world applications even with limited annotated data.
  
Building on these innovations, our architecture presents a novel hybrid search query mechanism that seamlessly integrates semantic and structural insights to retrieve the most contextually relevant causal graphs. Following this methodology, each news event is embedded into a vector space to capture its semantic meaning, while its underlying causal relationships consisting of causes, effects, and triggers are analyzed using graph-based structural metrics. By integrating cosine similarity with structural cues, it effectively identifies the most relevant causal events. It calculates a hybrid score for each event by combining the cosine similarity of its embedding with a binary structural score based on the counts of cause, effect and trigger nodes. Weighted coefficients ensure a balanced contribution of these factors, allowing the retrieval of events that are both semantically relevant and structurally connected. Events that exceed a predefined threshold are ranked, and the top candidates serve as few-shot learning examples. This dual-layered retrieval process plays a crucial role in enhancing the accuracy and robustness of causal classification and tagging in news headlines.
  
As digital news continues to evolve rapidly, it presents a growing challenge for causal extraction due to complex media narratives. The motivation behind this work stems from conventional methods that often fail to capture implicit, multi-sentence causal relationships, which are crucial to understand the deeper context of news events. This becomes even more challenging due to the fluid nature of news reporting, where causal connections are often spread across multiple sentences or hidden within the story. To address these limitations, our work integrates graph-based causal reasoning, few-shot learning, and a hybrid search query mechanism utilizing both semantic and structural insights to enhance the accuracy and scalability of causal annotations in news headlines for a deeper understanding of event relationships. Ultimately, our architecture strengthens causal analysis such as informed decision-making and policy analysis, meeting a critical need in today’s fast-changing media landscape. In order to precisely detect multi-causal and causally dependent events onto each other, we proposed GraphRAG-Causal which includes graph-augmented prompt engineering techniques using XML based prompting to get proper contextual cues. To further strengthen causal news classification, our architecture embeds a few-shot learning strategy combined with graph retrieval mechanism. This research contributes to the field of causal reasoning in digital news analysis. This document details our comprehensive exploration of digital news causal inference, including methodology, experimental evaluation, and implications for future research. 
\\
The rest of the paper is structured as follows. \hyperref[sec:related]{Section 2} presents the related work done on causality from Pearl's fundamental causal principles to GraphRAG approaches. \hyperref[sec:method]{Section 3} introduces our methodology and how the 3 stages of our architecture work this includes Data Preparation, Graph retrieval, Hybrid Cypher Query, and LLM Inference/Visualization. \hyperref[sec:exp]{Section 4} presents the experimental results. Finally, \hyperref[sec:future]{Section 5} offers conclusions about the proposed work and future research directions.

\section{Related Work}
In this section, we survey a broad range of literature that lays the foundation for understanding and advancing causal inference in data-driven environments. We begin by examining the core principles of causality and the challenges of discerning genuine causal relationships from mere correlations in high-dimensional settings, as outlined in the subsection on Causality and AI: Foundational Concepts and Advanced LLM Capabilities. Next, we review how natural language processing and news analysis have been harnessed for causal inference in Causality in NLP: Exploring the Potential and Challenges of Causal Inference in Language Processing and News Causality: Integrated Analysis and Classification Approaches. We then shift focus to techniques that enhance model performance through retrieval-augmented generation, including graph-based enhancements discussed in Retrieval-Augmented Generation: Enhancing Large Language Models with Contextual Retrieval, Causality in RAG: Advancing Causal Reasoning through Graph-Augmented Retrieval Generation, and GraphRAG: Integrating Graph-Augmented Retrieval and Causal Reasoning for Enhanced LLM Performance. Finally, we synthesize these contributions by identifying key limitations and research gaps that motivate our proposed solution. A detailed evaluation of our approach is presented in \hyperref[sec:result]{Section 2}.

\label{sec:related}
\subsection{\textbf{Causality and AI: Foundational Concepts and Advanced LLM Capabilities}}
Recent studies \cite{ashwani} \cite{rawal} \cite{zhao} \cite{causality} have explored the causal reasoning capabilities of Large Language Models (LLMs), aiming to enhance their understanding of cause and effect. Judea Pearl's seminal work \cite{pearl} formalized causal relationships through graphical models, structural equations, and the do-calculus, enabling mathematical interventions in observational data and providing criteria (e.g., back-door and front-door adjustments) to isolate causal signals. Meanwhile, the survey by Guo et al. \cite{guo} evaluates how big data impacts causal inference, emphasizing challenges in high-dimensional settings and the need for frameworks to address confounding and causal learning in machine learning.
The emerging body of research on causality in natural language processing explores how language models can reason about cause-and-effect relationships, identify causal claims in text, and potentially incorporate causal structures into their architecture. This interdisciplinary work bridges traditional NLP techniques with causal inference frameworks, aiming to address fundamental limitations in how language models understand and represent causation beyond statistical correlation.
\\
\\
Ashwani et al. \cite{ashwani} propose the CARE-CA framework, which combines explicit causal detection using ConceptNet and counterfactual analysis with implicit detection through LLMs. This approach shows improved performance in causal reasoning tasks such as causal discovery and counterfactual reasoning, although its strength in enhancing explainability is offset by its complexity and the need for extensive data. A study by Rawal et al. \cite{rawal} evaluates LLMs like ChatGPT on causal discovery and counterfactual reasoning using diverse datasets including tabular data and images. Their research demonstrates that LLMs can generate causal graphs and provide counterfactual explanations, though performance varies with data complexity and prompt engineering requirements, highlighting both the potential and limitations of these models in causal inference. The paper Causal-CoG by Zhao et al. \cite{zhao} introduces a method to enhance multi-modal language models by focusing on causal relationships in context generation. This approach improves reasoning capabilities but may be constrained by the complexity of integrating multiple modalities. Judea Pearl’s seminal work on causality, as summarized by Schölkopf, B. et al. \cite{causality}, provides a foundational framework for understanding causal relationships in machine learning. Pearl's structural causal models (SCMs) offer a systematic approach to reasoning about causality using directed acyclic graphs (DAGs) and do-calculus to represent causal dependencies and perform causal inference yet integrating these models into AI systems remains complex and computationally demanding. The work of Asiaee et al. \cite{asiaee} explores how causal frameworks can enhance the reliability of AI systems by addressing issues such as fairness, privacy, and explainability, which are essential for trustworthy AI. Similarly, a study by Carloni et al. \cite{carloni} investigates the relationship between causality and explainable AI (XAI), identifying perspectives that critique XAI through a causality lens, utilize XAI for causality, and integrate causality into XAI. Their findings suggest that while incorporating causality into XAI can improve robustness and interpretability, it is also resource intensive. Additionally, Imai et al. \cite{imai} propose a method to enhance causal inference using generative AI, such as large language models (LLMs). Their approach demonstrates that LLMs can generate causal representations from unstructured data, such as text, which can then be used for causal effect estimation leveraging the internal representations of generated text to improve both accuracy and efficiency in causal inference.

\subsection{\textbf{Causality in NLP: Exploring the Potential and Challenges of Causal Inference in Language Processing}}
The papers \cite{feder} \cite{jin} \cite{yang} \cite{chen} \cite{doan} \cite{chen2023cheer} \cite{moreno2023fincausal} \cite{liu2023event} on causality in natural language processing (NLP) explore various aspects of causal inference, from foundational concepts to specific applications, each contributing unique insights while facing certain limitations. \\
A study by Feder, et al. (2021) \cite{feder} provides a comprehensive overview of causal inference in NLP, highlighting its importance in moving beyond predictive tasks to understand causal relationships. It discusses the challenges of estimating causal effects with text data and explores potential applications in improving model robustness, fairness, and interpretability. However, the paper also states that it is still difficult for us to evaluate such systems as unified definitions and benchmark datasets are still lacking, which limits the ability to advance the field.
The CausalNLP tutorial by Jin et al. (2022) \cite{jin} offers an accessible introduction to causal discovery and effect estimation for NLP practitioners. While it lays a solid foundation for understanding causal inference and its applications in NLP, the tutorial falls short of addressing the complexities and nuances of implementing causal inference in real-world tasks, which may leave some researchers puzzled. A survey by Yang, et al. (2022) \cite{yang} offers a detailed examination of methods of extracting causal relations from text, emphasizing the need for robust and accurate techniques. It highlights various approaches and their effectiveness in different contexts. However, the survey points out that existing methods often struggle with the complexity of causality in NLP, and this indicates a need for further research and development in this area. \\
The chapter "Causal Inference and Natural Language Processing" from the book "Machine Learning for Causal Inference" by Chen, et al. (2023) \cite{chen} explores the integration of NLP with causal inference, focusing on the statistical challenges and opportunities. The chapter discusses the potential of causal inference to enhance NLP applications with better bias mitigation and confounding. The chapter does provide most of the theoretical knowledge of how causal inference and NLP might integrate but does not explicitly provide guidelines for implementing causal inference in NLP, which could be a limitation for practitioners. A paper by Doan, et al. (2018) \cite{doan} demonstrates the application of NLP techniques to extract health-related causal information from social media data. It showcases potential of NLP in real-world scenarios, particularly in puclic health. However, this study is limited by the quality and representativeness of twitter data, and how to deal with noisy data. \\ CHEER by Chen, et al. (2023) \cite{chen2023cheer} the framework introduced a noval approach to identifying causal relationships between events in documents. The method leverages centrality measures and high-order interactions to improve the accuracy of event causality extraction. While the approach shows promise, it is computationally intensive and requires significant amount of resources for training and deployment. FinCausal by Moreno, et al. (2023) \cite{moreno2023fincausal}, this focused more on causal inference in financial documents, aiming to improve accuracy of causal relationship detection in this domain. This was a shared task that provided a platform for researchers to benchmark their methods against a common dataset. However, this was limited to only financial documents and the nature of those documents only and not generalized to other domains. Lastly ICE by Liu, et al. (2023) \cite{liu2023event} which proposed a framework for extracting causal relationships between events in text, emphasizing the importance of implicit interactions between cause and effect. This method introduces a template-based conditional generation approach and a knowledge distillation mechanism to enhance causal reasoning. While the approach does achieve state-of-the-art performance on specific datasets, the generalization to other domains remains to be tested.

\subsection{\textbf{News Causality: Integrated Analysis and Classification Approaches}}
Causality in news has been studied through various methodologies, including causal graph extraction, content analysis, and predictive modeling. Initial work by Radinsky et al. (2012) \cite{radinsky} proposed a causality learning framework in news events, using semantic NLP and LinkedData ontologies (20 billion relations) to predict future events which showcases its effectiveness in predicting real-world events. Subsequent studies have adopted diverse methodological approaches: Maisonnave et al., (2022) \cite{maisonnave} compared time-series causality methods, comparing methodologies such as Granger causality and PCMCI to find underlying dependencies in news data, reporting PCMCI to be less vulnerable to confounders. On the other hand, domain-specific studies, such as Peng et al. (2021) \cite{peng}, employ qualitative content analysis to investigate causality framing in cancer news, exposing media biases that favor lifestyle factors over systemic issues. In misinformation analysis,  Cheng et al. (2021) \cite{cheng} proposed a causal model based on a neural network to analyze the spread of fake news, adding user embeddings and propensity scores to counteract social media data bias. Parallel to such efforts, Balashankar et al. (2019) \cite{balashankar} built the Predictive Causal Graph (PCG) framework that identified latent relationships among news events using temporal word co-occurrences and maintaining stock price prediction error rates between 1.5\% and 5\% over four years. Despite these advancements, it still remains a problem to work with implicit causality and real-time streams.
\\
Various methodologies have been proposed in the field of causal news classification to address the challenges of identifying causal relations in news text. The Causal News Corpus (CNC) by Tan, F. A., et al. (2022) \cite{tanNewsCorpus} presents a dataset of protest event news sentences annotated for causality and evaluates a BERT-based model achieving an 81.2\% F1-score. While CNC enhances causal relation detection, it lacks explicit cause-effect tagging and generalization beyond protest-related news. In the financial domain, the study by Wan et al. (2022) \cite{wan2022financial} explores causal sentence recognition, addressing complex patterns like multiple causes and effects. It proposes a BERT-CNN model that integrates local text representation with self-attention for improved classification. Experimental results show a 5.31-point F1-score improvement over previous models but it is limited to financial data. Chen et al. (2023) \cite{chen2023causal} propose the CCD framework for multi-modal fake news detection  utilizing causal intervention through backdoor adjustment to remove psycholinguistic bias from text features and counterfactual reasoning to reduce image-only bias by integrating these strategies with fusion-based models. However, the approach is limited by its reliance on fixed confounder representations (e.g., LIWC-based categories) and exhibits data-dependent performance varying across datasets, implying  sensitivity to data-specific characteristics and potential computational overhead. MLModeler5 @ Causal News Corpus 2023 by Bhatia et al. \cite{bhatia} investigates using RoBERTa to classify causal events in protest-related socio-political news, experimenting with various text preprocessing strategies. The study finds that including stopwords improves causal detection (F1-score of 0.77437), yet the model's performance lags behind the baseline and is limited by its focus on binary classification and protest news. Aziz et al. \cite{aziz2022csecu} proposed a single framework for causal event classification by fine-tuning two transformer models, RoBERTa and its Twitter-specific variant, and aggregating their prediction scores with a weighted arithmetic mean. The fusion strategy maximizes performance by ranking at the top of the shared task. Its fixed weighting parameters reduce adaptability, and it may struggle with capturing complex causal dependencies in diverse textual contexts. Krumbiegel et al. \cite{krumbiegel2022nlp4itf} proposed an event causality classification framework that fine-tunes pre-trained models (RoBERTa and BERT) augmented with linguistic features like NER and cause-effect-signal spans. While this approach displays encouraging gains at development time, its improvements on diverse test sets is marginal, revealing generalizability challenges. 
\\
\\
HeadlineCause by Gusev et al. \cite{gusev2021headlinecause} provides a corpus of news headline pairs annotated for implicit causal relationships, with more than 5,000 English and 9,000 Russian headline pairs annotated through crowdsourcing. The corpus tries to overcome the difficulty of implicit causal relation detection in short text, which calls for both common sense and world knowledge. The short context available in headlines could limit the power to recognize complex causal relations, and crowdsourcing might bring about annotation noise. ARGUABLY by Kohli et al. \cite{kohli2022arguably} proposes an approach towards event causality identification through contextually augmented language models, specifically by applying sentence-level data augmentation with distillBERT embeddings and fine-tuning using DeBERTa and RoBERTa models that have been enhanced with cross-attention achieving F1 score of 0.8610; however, the Causal News Corpus 
is relatively small in size. NoisyAnnot by Nguyen et al. \cite{nguyen2022noisyannot} uses transformer-based models with specialized loss functions involving inter-annotator agreement and number of annotators to identify causality in news text, achieving an F1 score of 0.8501.

\subsection{\textbf{Retrieval-Augmented Generation: Enriching Contextual Retrieval with Graph-Enhanced Causal Reasoning in LLMs}}
Retrieval Augmented Generation (RAG) has emerged as a pivotal technique for enhancing the capabilities of Large Language Models (LLMs) by integrating external knowledge base retrieval process into the generation which provides more context to the question/input from user. A survey by Gao, et al. \cite{gao2023retrieval} provides a comprehensive overview of RAG, detailing its architecture, applications, and advancements. RAG combines the strength of retrieval based and generation based models, allowing LLMs to access and retrieve relevant data from external sources during generation. This approach significantly improves the factual accuracy and contextuality  of generated content, addressing the limitations of standalone generative models that often lack up-to-date or domain specific knowledge. The survey demonstrates various RAG implementations, including dense retrieval methods and hybrid models, highlighting the effectiveness in tasks such as question answering, summarization, and dialogue systems. However, the survey also states the limitations such as computational cost of retrieval and the need of high quality datasets, diverse datasets to prevent redundancy and enhance generalization. Overall, RAG has a promising architecture for advancing LLMs, and offers a scalable solution to bridge the gap between model generated text and real world information.
\\
The paper by Samarajeewa, et al. \cite{samara} introduces a noval approach to enhancing causal reasoning in large language models (LLMs) by integrating causal graph RAG. This method aims to improve model's capability of understanding and generating text involving complex causal relationships by leveraging causal graphs created through a fine tuned version of BART for cause effect extraction to provide additional context and structure to the LLMs. The authors propose a framework that combines retrieval based methods with generative models, allowing LLMs to utilize relevant external knowledge during text generation. This integration substantially enhances the model's understanding of causal relationships, improving the accuracy and coherence of the generated text. The study shows effectiveness in various tasks such as question answering and text generation, showing notable improvements in causal reasoning capabilities. However, the paper also highlights the need of a high quality and diverse dataset for better understanding of causal inference for the LLMs. Additionally, the usage of BART, which was already fine tuned for cause-effect extraction, raises questions about the accuracy of BART itself for causality extraction. Overall, the research underlines the potential of causal graphs to advance causal reasoning in LLMs while also revealing areas for further development.

\subsection{\textbf{GraphRAG: Integrating Graph-Augmented Retrieval and Causal Reasoning for Enhanced LLM Performance}}
After advancements of LLMs with RAG there is another study which uses graph databases as external knowledge base for retrieval. GraphRAG represents a significant advancement in the field of AI by integrating graph-based knowledge representation with retrieval augmented generation (RAG) techniques. This approach enhances the ability of Large Language Models (LLMs) to understand and generate contextually relevant and accurate responses. By leveraging knowledge graphs, GraphRAG provides a more structured and semantically rich retrieval process, enabling LLMs to access and utilize information in a meaningful way. The paper by Han, et al. \cite{han2025graphrag} provides a comprehensive overview of this approach, highlighting its key components and techniques. GraphRAG integrates graph structured data into the retrieval augmented generation framework, allowing for more effective and contextually rich information retrieval. The authors propose a holistic GraphRAG framework that includes a query processor, retriever, organizer, generator and data source. They also reviewed techniques tailored to specific domains, emphasizing the unique challenges and opportunities presented by graph-structured data. However, the paper also acknowledges the challenges associated with implementation of GraphRAG. The complexity of graph construction and computational demands of processing large and complex graphs pose significant hurdles.Additionally, the need of high quality diverse datasets are needed to train the models effectively remains a critical flaw. Despite these limitations, GraphRAG offers a promising approach for advancing information retrieval and NLP.
\\
\\
The integration of causal reasoning into Graph Augmented Large Language Models (LLMs) has emerged as a promising approach to enhance complex reasoning capabilities. The paper by Luo, \cite{luo2025causal} explores the use of causal graphs to improve the reasoning abilities of LLMs. This approach leverages the structured nature of causal graphs to provide a more comprehensive understanding of relationships and dependencies, enabling LLMs to generate more accurate and contextually relevant responses. The study demonstrated effectiveness in tasks such as question answering and text generation, showing significant improvements inn handling complex queries. However, there were some limitations to highlight such as uncertainty in Chain-Of-Thought generation, the reliance on COT steps introduces variability, as different runs may produce inconsistent intermediate states. This stochasticity can affect the retrieval success for the same query and can fail. Another limitation is that the method relies on a causal sub-graph derived from a large knowledge graph (e.g., SemMedDB) and missing edges or over reliance on correlational links can dilute reasoning paths, forcing fallback to less relevant data. This also has a high dependency on High-Quality LLMs, its performance is bounded by the underlying LLM's capacity. Smaller or less advanced models may struggle to generate coherent CoTs or leverage retrieved paths effectively.

\subsection{\textbf{Limitations of Existing Approaches and Research Gaps}}
Existing research on causality spans on foundational theories, AI applications, and NLP specific challenges, yet notable limitations persist. Judea Pearl's work formalizes causal inference using graphical models and do-calculus, but integrating these frameworks into AI systems remains computationally extensive and very complex \cite{pearl, causality}. In AI, while studies like CARE-CA enhance explainability using causal inference, they require extensive data and complex integrations \cite{ashwani}. LLMs show potential in causal reasoning stated in \cite{rawal} but have limitations due to prompt engineering variability and domain specific constraints \cite{zhao, imai}. NLP focused work highlights the need for robust causal extraction methods, yet these struggle with data complexity, benchmarks, evaluation and implementation challenges \cite{feder, jin, yang}. Financial and event specific studies such as FinCausal and CHEER, advance in their own domains but hinder in generalization \cite{moreno2023fincausal, chen2023causal}. Causal news classification methods like CNC and CCD, improve detection but are narrowed down to domain specific datasets and reliance on fixed confounder \cite{tanNewsCorpus, chen2023causal}. RAG enhances LLMs with external knowledge but has extensive computational costs and data diversity issues \cite{gao2023retrieval}. Causal Graph RAG and GraphRAG introduce graph-based retrieval, yet they face challenges in graph construction complexity and scalability \cite{samara, han2025graphrag}. Causal Inference in GraphRAG as discussed in paper by Luo, \cite{luo2025causal} has limitations such as need of high quality LLMs and reliance on COT steps. Our approach GraphRAG-Causal proposes a hybrid retrieval architecture which bridges critical gaps in existing approaches, By combining cosine similarity and graph based searches, we reduce redundancy on large, high quality datasets, enabling smaller models to achieve competitive performance in causal news classification. The hybrid approach also mitigates the computational demands of graph construction and retrieval, making causal reasoning feasible for real-time applications. Unlike domain specific methods (e.g., FinCausal \cite{moreno2023fincausal}), our architecture leverages generic causal graphs and embeddings, enhancing adaptability to diverse contexts, the approach can be shifted to any domain with small amount of data. Unlike many methods relying solely on LLM generated COT steps, our hybrid query stabilizes retrieval, reducing stochasticity and improving consistency. By embedding causal graphs with tagged cause-effect relationships, we address limitations in detecting implicit causal links in short text (e.g., news headlines). This end-to-end pipeline integrates structured graph knowledge with LLMs, offering a scalable, efficient solution for causal news classification while addressing key limitations of prior work.

\begin{figure*}[t] 
    \includegraphics[width=\linewidth]{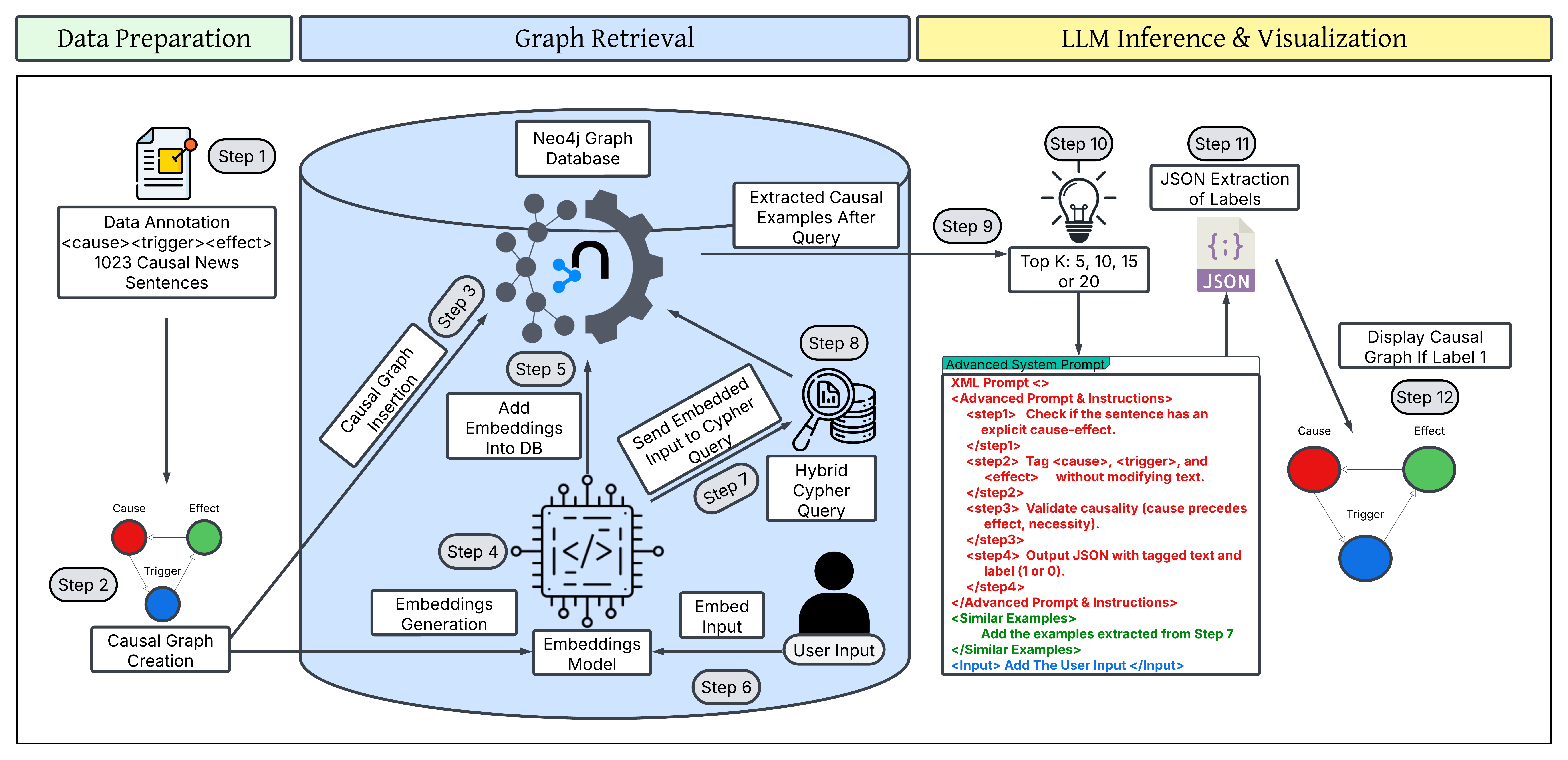}
    \caption{Overview of the proposed approach, illustrating a three-stage pipeline: (1) Data Preparation (data annotation and ingestion into a causal graph), (2) Graph Retrieval (querying the Neo4j database and leveraging embeddings for relevant context), and (3) LLM Inference and Visualization (using advanced system prompts for causal reasoning and displaying labeled outputs). This end-to-end workflow integrates graph-based knowledge with large language models to facilitate structured insights and intuitive visualizations.}
    \label{fig:enter-label}
\end{figure*}

\section{Methodology}
\label{sec:method}
In Graph Retrieval-Augmented Generation, NER (Named Entity Recognition) is applied to external knowledge documents and further converted into knowledge graphs which contain graph knowledge of each entity in the document. This approach is mostly done by using an LLM to create knowledge graphs from external documents and further create a cypher query using the same LLM to retrieve answers to advanced questions. In our approach, our aim is to have causal news headlines with tagged information converted into causal graphs (cause, effect, trigger) and embeddings to be inserted into a graph database and retrieved through a hybrid query which contains both cosine similarity and graph based similarity search. During the retrieval phase, the hybrid query is used to retrieve causal graphs, which are then provided to the LLM for few-shot learning and for leveraging external knowledge to classify and tag the news headlines. We have three primary stages: Data Preparation, Graph Retrieval and LLM Inference with visualizations of causal graph.

\subsection{\textbf{Data Preparation}}
The first involves acquiring news sentences from Causal News Corpus \cite{tanNewsCorpus} and start manually annotating 1023 sentences, this means tagging the sentences with cause, effect and trigger tags. Cause tag refers to the cause of the event that happened, effect is the event that happened because of the cause and trigger is like a signal which helps us in classifying whether the sentence is causal or not, triggers are mostly words that are leading to a cause or effect such as 'due to', 'because', etc. 
These annotated sentences are further processed and converted into causal graphs which have nodes and edges representing cause, effect and triggers. These graphs are then inserted into the graph database with the relationships. Here is a detailed analysis of the dataset that is inserted:

\begin{itemize}
    \item Total sentences: 2005 (acquired from Causal News Corpus)
    \item Tagged sentences: 1030
    \item Events: 1021 (extracted from 1030 which were relevant)
    \item Causes: 1147
    \item Effects: 1118
    \item Triggers: 1102
    \item Relationships: 3404 
    \item Three relationship types: 
    \item \textit{CAUSES}
    \item \textit{RESULTS\_IN}
    \item \textit{HAS\_TRIGGER}
\end{itemize}

\subsection{\textbf{Graph Retrieval}}
In second stage, we have the main components of graph RAG. After data preparation, we inserted the causal graphs into our graph database. Our approach is different than the usual graph RAG where we are also adding embeddings for the causal graphs into our graph database. The embeddings are for the events and each event will have its own embedding vector stored in its properties in the graph database. The algorithm for adding embeddings is provided in Algorithm 1.

\begin{algorithm}
\caption{Graph Embeddings Update in GraphDB}
\begin{algorithmic}[0]
\State \textbf{1. Clean Existing Embeddings}
\begin{itemize}
\item Retrieve all nodes of type \textit{Event}, \textit{Cause}, \textit{Effect}, and \textit{Trigger}.
\item Set their \texttt{embedding} attribute to \texttt{NULL}.
\end{itemize}
\State \textbf{2. Recreate Vector Indexes}
\begin{itemize}
\item For each node type (\textit{Event}, \textit{Cause}, etc.):
\begin{itemize}
\item Check if an existing vector index exists and delete it if found.
\item Create a new vector index with the following parameters:
\begin{itemize}
\item \textbf{Dimension}: 384
\item \textbf{Similarity metric}: Cosine similarity
\end{itemize}
\end{itemize}
\end{itemize}
\State \textbf{3. Generate Embeddings}
\begin{itemize}
\item Extract nodes of type \textit{Event}, \textit{Cause}, \textit{Effect}, or \textit{Trigger} that have non-null text.
\item Batch processing:
\begin{itemize}
\item Divide retrieved nodes into batches of size b.
\item For each batch:
\begin{itemize}
\item Extract text from each node.
\item Use a pre-trained embedding model to compute embeddings for the batch.
\item Update each node in the batch by setting its \texttt{embedding} attribute to the generated embedding.
\end{itemize}
\end{itemize}
\end{itemize}
\State \textbf{4. Verify Embedding Creation}
\begin{itemize}
\item Query the graph to count nodes of each type (\textit{Event}, \textit{Cause}, etc.) and determine how many have valid embeddings.
\item Report results listing each node type, total count, and the number of nodes with embeddings.
\end{itemize}
\end{algorithmic}
\end{algorithm}

The embeddings are inserted using cosine similarity to find which embedding aligns with the appropriate index (steps 4 and 5). After embeddings, we will take user input, embed the input, and further send embedded input to cypher query for retrieval of semantically similar news sentences. Here in step 8 we are performing a hybrid search query which works on both cosine similarity and structural scores.

\subsection{\textbf{Hybrid Cypher Query - Mathematical Formulations}}

Let $\mathcal{E}$ denote the set of all events in the graph database. For each event $e \in \mathcal{E}$, we denote its embedding vector by $\mathbf{e} \in \mathbb{R}^n$ and its associated text by $\text{text}(e)$. Furthermore, let $\mathbf{q} \in \mathbb{R}^n$ be the query embedding. We restrict our attention to events with non-null embeddings and text, i.e.,
\begin{equation}
\mathcal{E}' = \{ e \in \mathcal{E} \mid \mathbf{e} \neq \text{NULL} \text{ and } \text{text}(e) \neq \text{NULL} \}.
\end{equation}

For each event $e \in \mathcal{E}'$, we define three counts based on its causal relationships in the graph. The \emph{Effects Count} is given by
\begin{equation}
N_{\text{effect}}(e) = \left| \{ x \mid (e \xrightarrow{\text{RESULTS\_IN}} x) \} \right|,
\end{equation}
where $x$ is an \textit{Effect} node. Similarly, the \emph{Causes Count} is defined as
\begin{equation}
N_{\text{cause}}(e) = \left| \{ x \mid (x \xrightarrow{\text{CAUSES}} e) \} \right|,
\end{equation}
with $x$ representing a \textit{Cause} node, and the \emph{Triggers Count} is defined as
\begin{equation}
N_{\text{trigger}}(e) = \left| \{ x \mid (e \xrightarrow{\text{HAS\_TRIGGER}} x) \} \right|,
\end{equation}
where $x$ is a \textit{Trigger} node.

To capture whether an event exhibits any causal connections, we introduce a \emph{structural score} $S(e)$ defined as
\begin{equation}
S(e) =
\begin{cases}
1, & \text{if } N_{\text{effect}}(e) + N_{\text{cause}}(e) + N_{\text{trigger}}(e) > 0, \\
0, & \text{otherwise}.
\end{cases}
\end{equation}
Equivalently, using the indicator function $\mathbb{I}\{\cdot\}$, we can express this as
\begin{equation}
S(e) = \mathbb{I}\{ N_{\text{effect}}(e) + N_{\text{cause}}(e) + N_{\text{trigger}}(e) > 0 \}.
\end{equation}

In addition, we quantify the semantic similarity between an event $e$ and the query using cosine similarity:
\begin{equation}
\text{sim}(e, \mathbf{q}) = \frac{\mathbf{e} \cdot \mathbf{q}}{\|\mathbf{e}\| \, \|\mathbf{q}\|}.
\end{equation}
Although cosine similarity generally ranges between $-1$ and $1$, our embeddings are normalized so that higher values indicate greater similarity.

To integrate the semantic similarity with the structural information, we define a \emph{hybrid score} $H(e)$ for each event as follows:
\begin{equation}
H(e) = \alpha \, \text{sim}(e, \mathbf{q}) + \beta \, S(e),
\end{equation}
where $\alpha \in \mathbb{R}^+$ and $\beta \in \mathbb{R}^+$ are weights that balance the contributions of the embedding similarity and the structural cue, respectively.

We then apply a filtering step by selecting only those events for which
\begin{equation}
H(e) \ge \tau,
\end{equation}
with $\tau$ being a predefined threshold. The events satisfying this condition are ranked in descending order based on their hybrid scores, and the top $k$ events are selected:
\begin{equation}
\text{Result} = \operatorname{Top}_k\left( \{ e \in \mathcal{E}' \mid H(e) \ge \tau \} \right).
\end{equation}

Finally, for each event $e$ in the result set, we retrieve the associated textual information from its connected nodes. Specifically, we define:
\begin{align}
T_{\text{effect}}(e) &= \{ \text{text}(x) \mid (e \xrightarrow{\text{RESULTS\_IN}} x), \, x \in \text{EffectNodes} \}, \\
T_{\text{cause}}(e)  &= \{ \text{text}(x) \mid (x \xrightarrow{\text{CAUSES}} e), \, x \in \text{CauseNodes} \}, \\
T_{\text{trigger}}(e) &= \{ \text{text}(x) \mid (e \xrightarrow{\text{HAS\_TRIGGER}} x), \, x \in \text{TriggerNodes} \}.
\end{align}

In summary, the overall hybrid scoring function for an event $e$ is given by
\begin{equation}
H(e) = \alpha \left( \frac{\mathbf{e} \cdot \mathbf{q}}{\|\mathbf{e}\| \, \|\mathbf{q}\|} \right) + \beta \, \mathbb{I}\{ N_{\text{effect}}(e) + N_{\text{cause}}(e) + N_{\text{trigger}}(e) > 0 \}.
\end{equation}
Events with $H(e) \ge \tau$ are ranked in descending order, and the top $k$ events, along with their associated causal texts, are returned for further analysis.

The hybrid query search amplifies the chances of acquiring semantically similar sentences to the input text given by the user. The Cypher query in listing 1 shows all the score calculations required for the final top k selected examples for few shot learning of our LLM (steps 6, 7, and 8). 

\subsection{\textbf{LLM Inference and Visualization}}
Step 9: Few-Shot Learning Integration after Knowledge Extraction: \\
After ranking and achieving the top-k events by their hybrid scores, we select these top-k events which are the most relevant to the input causal news event. These top-k examples are then incorporated into the XML prompt as few-shot learning examples for classification and tagging. This integration provides the LLM with concrete instances and solid examples, helping it to better understand the task and how to detect causal relationships.

Step 10: XML-Based Prompting: \\
In this step, we employ an XML-formatted prompt to deliver detailed instructions to the LLM this includes the 5 causality tests from Causal News Corpus \cite{tanNewsCorpus}. 

\begin{lstlisting}[caption={Cypher Query for Causal Graph Retrieval}, label={lst:cypher_query}]
MATCH (e:Event)
    WHERE e.embedding IS NOT NULL AND e.text IS NOT NULL

    // Efficient connection counting using COUNT
    OPTIONAL MATCH (e)-[:RESULTS_IN]->(effect:Effect)
    WITH e, COUNT(effect) AS effect_count

    OPTIONAL MATCH (e)<-[:CAUSES]-(cause:Cause)
    WITH e, effect_count, COUNT(cause) AS cause_count

    OPTIONAL MATCH (e)-[:HAS_TRIGGER]->(trigger:Trigger)
    WITH e, effect_count, cause_count, COUNT(trigger) AS trigger_count

    // Calculate structural score (maintain original binary approach)
    WITH e, effect_count, cause_count, trigger_count,
         CASE 
             WHEN effect_count + cause_count + trigger_count = 0 
             THEN 0.0 
             ELSE 1.0 
         END AS structural_score

    // Calculate similarity and hybrid score
    WITH e, effect_count, cause_count, trigger_count, structural_score,
         gds.similarity.cosine(e.embedding, $query_embedding) AS embedding_similarity,
         ($embedding_weight * gds.similarity.cosine(e.embedding, $query_embedding) + $structure_weight * structural_score) AS hybrid_score

    WHERE hybrid_score >= $similarity_threshold

    // Late text collection for final candidates
    CALL {
        WITH e
        OPTIONAL MATCH (e)-[:RESULTS_IN]->(effect:Effect)
        RETURN COLLECT(effect.text) AS effect_texts
    }
    CALL {
        WITH e
        OPTIONAL MATCH (e)<-[:CAUSES]-(cause:Cause)
        RETURN COLLECT(cause.text) AS cause_texts
    }
    CALL {
        WITH e
        OPTIONAL MATCH (e)-[:HAS_TRIGGER]->(trigger:Trigger)
        RETURN COLLECT(trigger.text) AS trigger_texts
    }

    RETURN e.text AS text,
           e.id AS event_id,
           hybrid_score,
           embedding_similarity,
           structural_score,
           effect_count,
           cause_count,
           trigger_count,
           effect_texts,
           cause_texts,
           trigger_texts
    ORDER BY hybrid_score DESC
    LIMIT $top_k
\end{lstlisting}

The XML prompt not only includes the top-k few-shot examples from Step 9 but also outlines specific directives and rules the LLM has to follow. The last instruction is to explicitly output a JSON object which has two keys: 1: tagged\_sentence and 2: label.
\\
These strict rules and instructions in XML ensure that LLM adheres to the required format and delivers consistent, reliable results which is proved later in experiments. Steps 11 and 12 are just extraction of JSON object and if label is 1 display causal graph which is similar to step 2. This approach combining XML prompting with few-shot learning has proven to be more effective than normal prompt engineering with similar results to fine tuning models.

\section{Experimental Evaluation}
\label{sec:exp}
This section presents the evaluation of the proposed approach using a test dataset from Causal News Corpus \cite{tanNewsCorpus} which included gold labels from crowd sourcing. The metrics used are F1, Recall, Accuracy, Precision and Matthews Correlation Coefficient scores. For the causal graphs creation 1030 sentences were annotated but 1021 were selected as gold annotated sentences for insertion in graph database. Neo4j \cite{neo4j} graph database was used as the knowledge base. Huggingface embedding model (all-MiniLM-L6-v2) \cite{huggingface} was used, which is trained on multiple text corpora that produce 384-dimensional vector embeddings for a given text was used. Deepseek's model deepseek-r1-distill-llama-70b was used through Groq \cite{groq} API. The results were conducted using a fixed context window size with 6k tokens per minute maximum length (rank upto 20 max examples for few shot learning). Test dataset was acquired from Causal News Corpus \cite{tanNewsCorpus} which contains gold labels with agreement scores. The tests were performed on top 5, 10, 15 and 20 examples with increasing evaluation scores in each knowledge base increase for LLM.

\subsection{\textbf{Hybrid Query Mechanism and Top-Retrieved Event Analysis}}

Our hybrid query approach integrates semantic and structural analyses to retrieve events that closely match a given query sentence. In this framework, the query is processed along two parallel paths. First, the sentence is converted into an embedding representation that captures its semantic meaning. Simultaneously, a structural analysis extracts features that describe the narrative pattern and causal relationships. The two scores are then combined to form a single hybrid score used for ranking.

\begin{figure}[h]
    \centering
    \begin{tikzpicture}[node distance=2cm, auto]
        % Define blocks
        \node (query) [rectangle, draw, rounded corners, align=center, minimum width=2.8cm, minimum height=1cm] {Input Query Sentence};
        \node (embedding) [rectangle, draw, rounded corners, below left of=query, xshift=-1.5cm, minimum width=2.8cm, minimum height=1cm] {Embedding Generation};
        \node (structural) [rectangle, draw, rounded corners, below right of=query, xshift=1.5cm, minimum width=2.8cm, minimum height=1cm] {Structural Analysis};
        \node (combine) [rectangle, draw, rounded corners, below of=query, yshift=-3cm, minimum width=3.5cm, minimum height=1cm, align=center] {Hybrid Score Calculation\\ (Embedding + Structural)};
        \node (retrieve) [rectangle, draw, rounded corners, below of=combine, yshift=-1cm, minimum width=2.8cm, minimum height=1cm] {Event Retrieval};
        
        % Draw arrows
        \draw[->] (query) -- (embedding);
        \draw[->] (query) -- (structural);
        \draw[->] (embedding) -- (combine);
        \draw[->] (structural) -- (combine);
        \draw[->] (combine) -- (retrieve);
    \end{tikzpicture}
    \caption{Workflow of the Hybrid Query Mechanism.}
    \label{fig:workflow}
\end{figure}
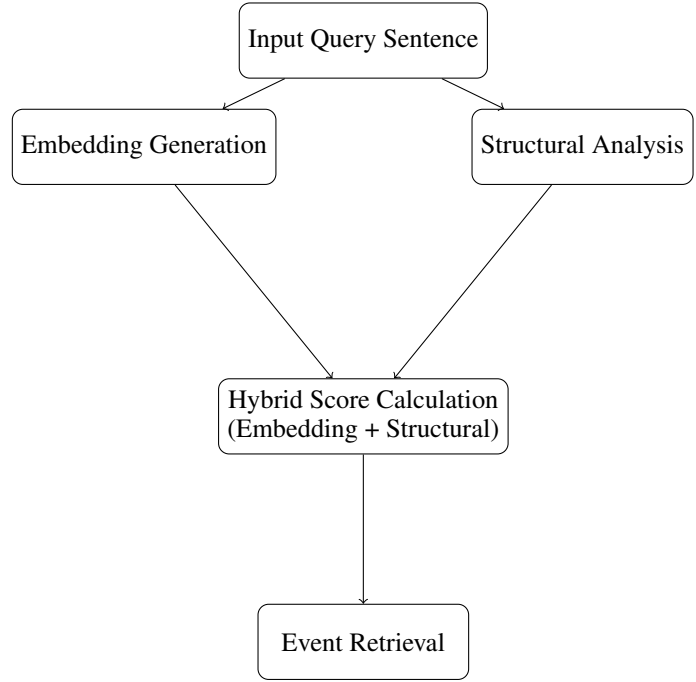

Figure~\ref{fig:workflow} illustrates the overall workflow of this approach. The query sentence is first encoded into an embedding and analyzed for structural components. The resulting scores are then merged, and events are ranked by their final hybrid score.

\begin{figure}[h]
    \centering
    \begin{tikzpicture}
    \begin{axis}[
        ybar,
        symbolic x coords={Event 1, Event 2, Event 3, Event 4, Event 5},
        xtick=data,
        ylabel={Hybrid Score},
        xlabel={Retrieved Events},
        nodes near coords,
        ymin=0.65, ymax=1.05,
        width=0.9\linewidth
    ]
    \addplot coordinates {(Event 1,1.0) (Event 2,0.757) (Event 3,0.743) (Event 4,0.684) (Event 5,0.678)};
    \end{axis}
    \end{tikzpicture}
    \caption{Hybrid Scores for the Retrieved Similar Events.}
    \label{fig:hybridBar}
\end{figure}
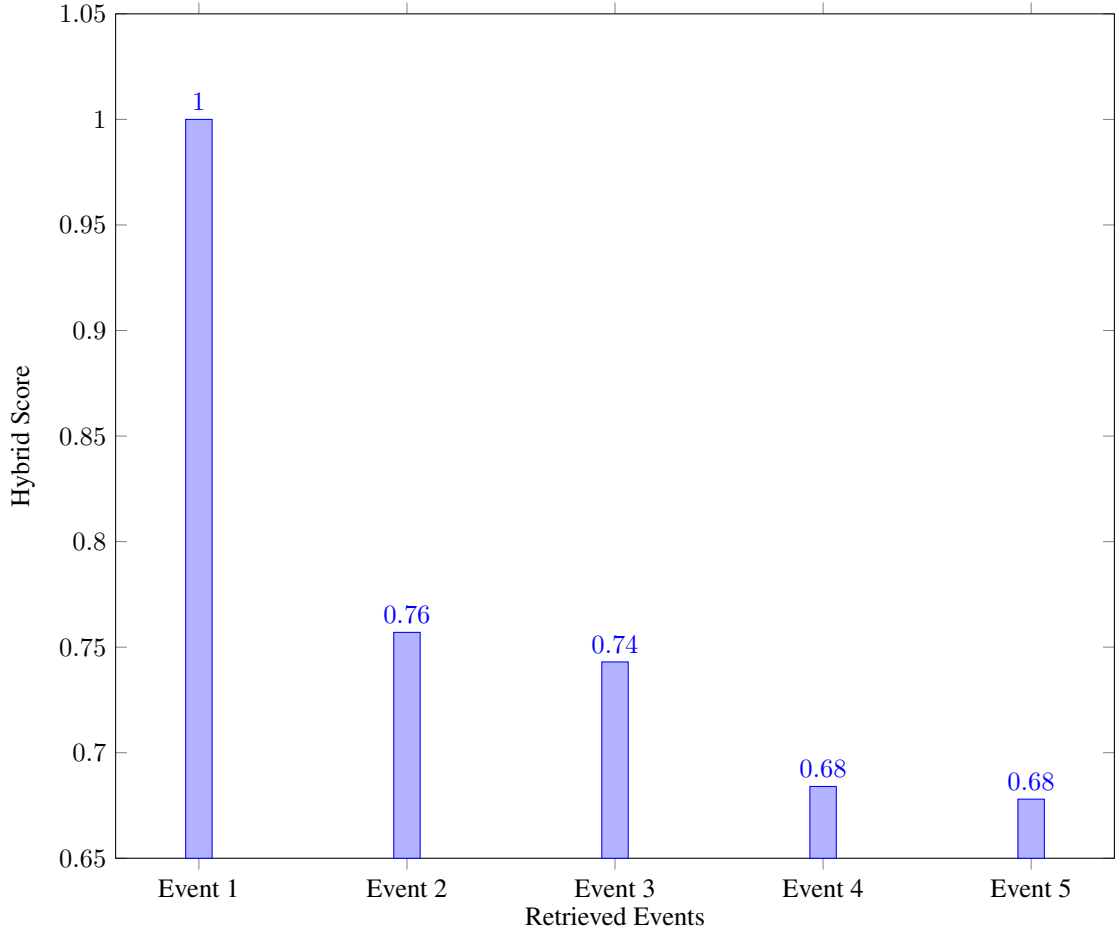

To demonstrate the approach, a hybrid query was executed for the sentence: 
\textit{\textbf{"I observed the attack on the police, I have no doubt about it, '' Modiba said during cross-examination.}}
This query retrieved five similar events. Figure~\ref{fig:hybridBar} shows a bar chart of the hybrid scores for all retrieved events.

\medskip

The top three events with the highest hybrid scores are analyzed in table 1.

\begin{table}[h]
\centering
\caption{Top Three Retrieved Events and Similarity Rationale}
\label{tab:topEvents}
\begin{tabularx}{\linewidth}{@{}l X X@{}}
\toprule
\textbf{Event No.} & \textbf{Event} & \textbf{Similarity Rationale} \\
\midrule
\textbf{Event 1} & 
i observed the attack on the police , i have no doubt about it , '' modiba said during cross-examination . & 
This event is identical to the query sentence, with a perfect hybrid score (1.0), embedding similarity (1.0), and structural score (1.0). Its exact match in both semantic content and structural pattern confirms its relevance. \\
\midrule
\textbf{Event 2} & 
thanks to the providential arrest of a terrorist in mumbai and his subsequent interrogation , the police are asserting with a considerable degree of confidence that the let planned and orchestrated the attacks that took the lives of more than 180 people . & 
Although the sentence differs in content, it shares a similar structural pattern by describing a police-related incident. The event emphasizes a reported assertion (police claiming responsibility based on an interrogation) that parallels the narrative style of the query. Its perfect structural score (1.0) compensates for a moderate embedding similarity (0.595). \\
\midrule
\textbf{Event 3} & 
mpofu said lt-col kaizer modiba was fabricating his evidence about the violence during the mineworkers ' strike at marikana , near rustenburg in the north west , because he had not been in a position to witness an attack . & 
This event mentions Modiba and involves a statement regarding evidence and an observed incident, aligning it with the query. Despite a moderate embedding similarity (0.572), its perfect structural score (1.0) and the inclusion of key narrative elements (e.g., reporting of an event and causal language) support its relevance. \\
\bottomrule
\end{tabularx}
\end{table}

\begin{table}[htbp]
\centering
\small
\caption{\textbf{Evaluation Metrics Comparison: DeepSeek R1 Distill LLaMA 70B (Top K) vs.\ Causal News Corpus}}
\label{tab:comparison}
\begin{tabular}{lccccc}
\toprule
\textbf{Top K} & \textbf{F1} & \textbf{Acc} & \textbf{Prec} & \textbf{Rec} & \textbf{MCC} \\
\midrule
\textbf{CNC} & \textbf{0.8347} & \textbf{0.8111} & \textbf{0.8063} & \textbf{0.8652} & \textbf{0.6172} \\
\midrule
Top K = 5  & 0.7774 & 0.7678 & 0.8239 & 0.7359 & 0.5401 \\
Top K = 10 & 0.8046 & 0.7895 & 0.8235 & 0.7865 & 0.5774 \\
Top K = 15 & 0.8184 & 0.7926 & 0.7906 & 0.8483 & 0.5792 \\
Top K = 20 & 0.8216 & 0.7957 & 0.7917 & 0.8539 & 0.5856 \\
Top K = 35 & 0.7979 & 0.7593 & 0.7414 & 0.8636 & 0.5138 \\
Top K = 50 & 0.8152 & 0.7854 & 0.7812 & 0.8522 & 0.5637 \\
\midrule
\textbf{Top K = 40} & \textbf{0.8288} & \textbf{0.8} & \textbf{0.7868} & \textbf{0.8757} & \textbf{0.5948} \\

\bottomrule
\end{tabular}
\end{table}

\begin{table}[htbp]
  \centering
  \small
  \setlength{\tabcolsep}{6pt}
  \renewcommand{\arraystretch}{1.0}
  \caption{%
    \textbf{Evaluation Metrics Comparison: LLaMA 4 Maverick 17B Instruct 128B (Top K) vs.\ Causal News Corpus}%
  }
  \label{tab:llama4comparison}
  \begin{tabular}{lccccc}
    \toprule
    \textbf{Top K} & \textbf{F1} & \textbf{Acc} & \textbf{Prec} & \textbf{Rec} & \textbf{MCC} \\
    \midrule
    \textbf{CNC} & \textbf{0.8347} & \textbf{0.8111} & \textbf{0.8063} & \textbf{0.8652} & \textbf{0.6172} \\
    \midrule
    Top K = 25  & 0.7834 & 0.7214 & 0.6839 & 0.9167 & 0.4515 \\
    Top K = 30  & 0.8108 & 0.7520 & 0.7219 & 0.9247 & 0.4972 \\
    Top K = 40  & 0.7950 & 0.7328 & 0.6995 & 0.9209 & 0.4660 \\
    Top K = 50  & 0.7851 & 0.7137 & 0.6749 & 0.9384 & 0.4393 \\
    Top K = 30 (It 3) & 0.8120 & 0.7816 & 0.7720 & 0.8563 & 0.5575 \\
    Top K = 30 (It 4)   & 0.7988 & 0.7413 & 0.7112 & 0.9110 & 0.4794 \\
    \midrule
    \textbf{Top K = 30 (It 2)} & \textbf{0.8152} & \textbf{0.7568} & \textbf{0.7202} & \textbf{0.9392} & \textbf{0.5141} \\
    \bottomrule
  \end{tabular}
\end{table}

In summary, while Event 1 is an exact match, Events 2 and 3 are retrieved based on their high structural consistency and shared semantic cues (such as involvement of the police and reference to Modiba). These results validate the effectiveness of our hybrid query mechanism in capturing both the semantic and structural nuances required for accurate event retrieval.

\subsection{\textbf{Classification Performance Metrics}}

\begin{figure}[h]
    \centering
    \begin{tikzpicture}
        \begin{axis}[
            xlabel={Top K Examples},
            ylabel={F1 Score},
            ymin=0.75, ymax=0.85,
            xtick={5,10,15,20, 25, 30, 40, 50},
            ytick={0.75,0.78,0.80,0.82,0.84},
            grid=major,
            width=8cm, height=6cm,
            mark size=3pt,
            legend pos=south east
        ]
            \addplot[color=blue, mark=o, thick] coordinates {
                (5, 0.7774)
                (10, 0.8046)
                (15, 0.8184)
                (20, 0.8216)
                (25, 0.7979)
                (30, 0.8119)
                (40, 0.8288)
                (50, 0.8152)
            };
            \addlegendentry{F1 Score}
        \end{axis}
    \end{tikzpicture}
    \caption{F1 Score Growth for DeepSeek-R1-Distill-LLaMA-70B}
    \label{fig:F1Score}
\end{figure}
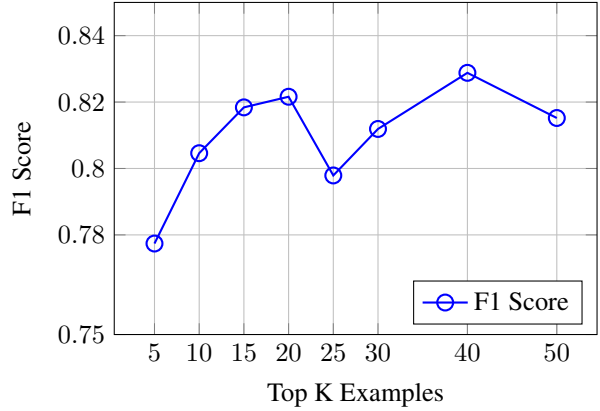

\begin{figure}[h]
    \centering
    \begin{tikzpicture}
        \begin{axis}[
            xlabel={Top K Examples},
            ylabel={F1 Score},
            ymin=0.75, ymax=0.85,
            xtick={25, 30, 40, 50},
            ytick={0.75, 0.78, 0.80, 0.82, 0.84},
            grid=major,
            width=8cm, height=6cm,
            mark size=3pt,
            legend pos=south east
        ]
            \addplot[color=red, mark=square*, thick] coordinates {
                (25, 0.7834)
                (30, 0.8108)
                (30, 0.8152) % Iteration 2 - highest F1
                (30, 0.8120) % Iteration 3 (>80% hyper)
                (30, 0.7988) % Iteration 4 (thresh 0.2)
                (40, 0.7950)
                (50, 0.7851)
            };
            \addlegendentry{F1 Score (Maverick)}
        \end{axis}
    \end{tikzpicture}
    \caption{F1 Score Growth for LLaMA 4 Maverick 17B Instruct 128B}
    \label{fig:F1ScoreMaverick}
\end{figure}
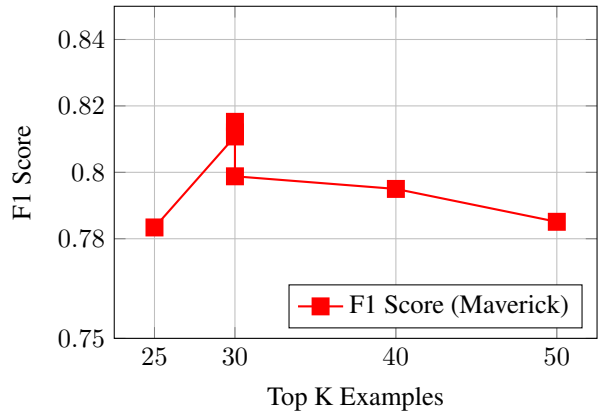

\begin{figure}[h]
    \centering
    \begin{tikzpicture}
        \begin{axis}[
            xlabel={Top K Examples},
            ylabel={Recall},
            ymin=0.85, ymax=0.95,
            xtick={25, 30, 40, 50},
            ytick={0.85, 0.88, 0.90, 0.92, 0.94},
            grid=major,
            width=10cm, height=7cm,
            mark size=3pt,
            legend style={at={(1.05,1)}, anchor=north west}
        ]

        % Main Run
        \addplot[color=orange, mark=*, thick] coordinates {
            (25, 0.9167)
            (30, 0.9247)
            (40, 0.9209)
            (50, 0.9384)
        };
        \addlegendentry{Main Run}

        % Iteration 2
        \addplot[color=blue, mark=square*, thick, dashed] coordinates {
            (30, 0.9392)
        };
        \addlegendentry{Iteration 2}

        % Iteration 3 - High Thresholds
        \addplot[color=purple, mark=triangle*, thick, dash dot] coordinates {
            (30, 0.8563)
        };
        \addlegendentry{Iteration 3 (Threshold $>$ 0.8)}

        % Iteration 4 - Low Thresholds
        \addplot[color=teal, mark=diamond*, thick, dotted] coordinates {
            (30, 0.9110)
        };
        \addlegendentry{Iteration 4 (Threshold = 0.2)}

        \end{axis}
    \end{tikzpicture}
    \caption{Recall Score Comparison for LLaMA 4 Maverick 17B Instruct 128B across Iterations}
    \label{fig:RecallMaverickIterations}
\end{figure}
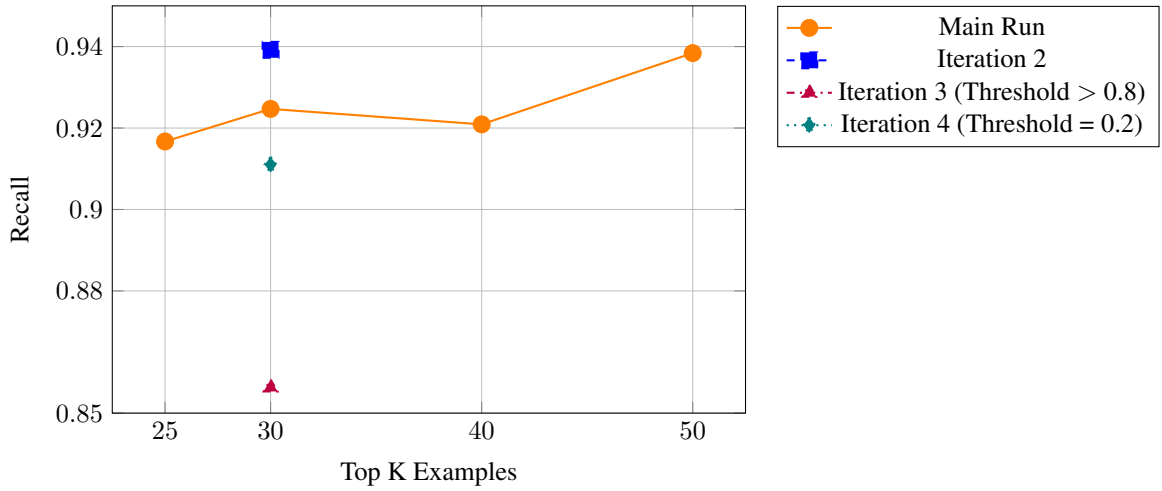

Table 2 illustrates a consistent increase in the F1 score with every five additional examples, demonstrating that as the knowledge base expands, the LLM achieves progressively better results. Notably, it can reach 90\% accuracy when provided with approximately 50 examples. Our approach highlights that LLMs perform more effectively when they have sufficient domain knowledge and are guided by well-defined instructions. For the tagging accuracy we are self evaluating the tagged sentences that are generated by the LLM.

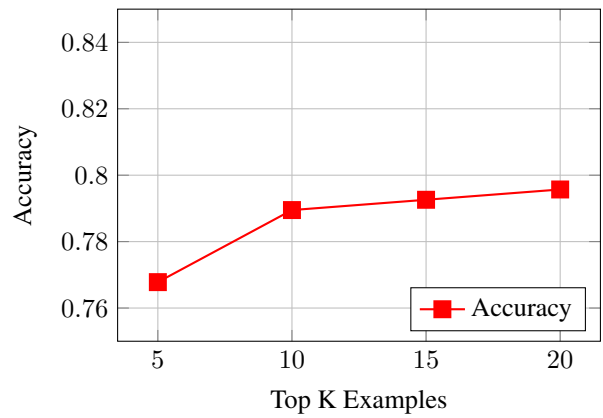
\begin{figure}[h]
    \centering
    \begin{tikzpicture}
        \begin{axis}[
            xlabel={Top K Examples},
            ylabel={Accuracy},
            ymin=0.75, ymax=0.85,
            xtick={5,10,15,20},
            ytick={0.76,0.78,0.80,0.82,0.84},
            grid=major,
            width=8cm, height=6cm,
            mark size=3pt,
            legend pos=south east
        ]
            \addplot[color=red, mark=square*, thick] coordinates {
                (5, 0.7678)
                (10, 0.7895)
                (15, 0.7926)
                (20, 0.7957)
            };
            \addlegendentry{Accuracy}
        \end{axis}
    \end{tikzpicture}
    \caption{Accuracy Trend for DeepSeek-R1-Distill-LLaMA-70B}
    \label{fig:Acc}
\end{figure}

\begin{figure}[h]
    \centering
    \begin{tikzpicture}
        \begin{axis}[
            xlabel={Top K Examples},
            ylabel={MCC},
            ymin=0.53, ymax=0.60,
            xtick={5,10,15,20},
            ytick={0.54,0.56,0.58,0.60},
            grid=major,
            width=8cm, height=6cm,
            mark size=3pt,
            legend pos=south east
        ]
            \addplot[color=orange, mark=star, thick] coordinates {
                (5, 0.5401)
                (10, 0.5774)
                (15, 0.5792)
                (20, 0.5856)
            };
            \addlegendentry{MCC}
        \end{axis}
    \end{tikzpicture}
    \caption{MCC Trend for DeepSeek-R1-Distill-LLaMA-70B}
    \label{fig:MCC}
\end{figure}
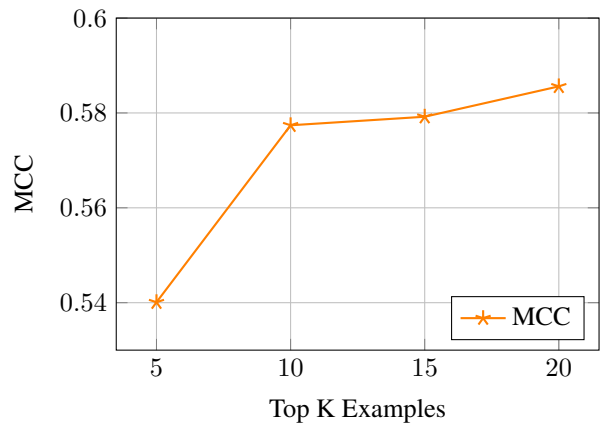

Figure ~\ref{fig:F1Score} presents the growth in the F1 score as the number of examples increases. In addition, Figures ~\ref{fig:Acc}, and ~\ref{fig:MCC} illustrate the trends in accuracy, precision and recall, and MCC, respectively. These graphical representations provide an intuitive understanding of the model's performance and further support the observation that additional domain-specific examples lead to improved classification results.

The graphical analysis reinforces the observations from Table 2. In particular, the accuracy and MCC metrics steadily improve with the increase in the number of examples, while the precision and recall metrics exhibit slight variations. These results underline the importance of expanding the domain-specific knowledge base and fine-tuning instructions to achieve enhanced model performance.

\section{Conclusion and Future Work}
\label{sec:future}
In conclusion, we present a novel approach to classification and tagging by leveraging Graph RAG rather than traditional fine-tuning methods. Our method effectively addresses both tagging and classification tasks, aided by a specialized knowledge base of causal graphs and relationships. The retrieval mechanism employs a hybrid strategy, distinct from purely LLM-based approaches, and incorporates a new Cypher query technique that has not been previously used in this context. A key strength of our approach lies in the scoring mechanism, which is further enhanced by an XML-based prompt that delivers improved performance. Experimental results show consistent F1 score gains, underscoring the effectiveness of our method.

For future work, we plan to develop a more comprehensive evaluation system (a Validation Layer) for tagging, as there is currently no standardized accuracy measure for this component. We also aim to integrate AI RAG Agents, enabling step-by-step data retrieval within larger LLMs for faster inference and fewer token constraints. This expansion will further advance our framework’s applicability and scalability in real-world scenarios.

\end{document}